\documentclass[fleqn,twoside]{article}
\usepackage{espcrc2}

\usepackage{graphicx}
\usepackage{epsfig}
\usepackage[figuresright]{rotating}

\def\greaterthansquiggle{\raise.3ex\hbox{$>$\kern-.75em\lower1ex
\hbox{$\sim$}}}
\def\lessthansquiggle{\raise.3ex\hbox{$<$\kern-.75em\lower1ex
\hbox{$\sim$}}}

\newcommand{\ci}{\cite}
\newcommand{\beqn}{\begin{eqnarray}}
\newcommand{\eeqn}{\end{eqnarray}}
\newcommand{\bequ}{\begin{equation}}
\newcommand{\eequ}{\end{equation}}
\newcommand{\bsl}{\begin{sloppypar}}
\newcommand{\esl}{\end{sloppypar}}

\newcommand{\AmS}{{\protect\the\textfont2
  A\kern-.1667em\lower.5ex\hbox{M}\kern-.125emS}}

\title{New physics searches at a Linear Collider with polarized beams }

\author{G. Moortgat-Pick\address[HH]{Inst. f\"ur Theoretische Physik, 
        Universit\"at Hamburg, D--22607 Hamburg, Germany }
        \thanks{Talk given by G. Moortgat-Pick, gudrid@mail.desy.de},
        A.Bartl\address[Wien]{Institut f\"ur Theoretische Physik, 
Universit\"at Wien, A-1090 Vienna, Austria}, 
        K. Hidaka\address[Jap]{Dept. of Physics, Tokyo Gakugei University, 
Koganei, Tokyo 184-8501, Japan}, 
T. Kernreiter\addressmark[Wien], 
H.Liivat\address[Tu]{Institute of Physics, University of Tartu, Riia 142, 
51014 Tartu, Estonia}, R.-K.Loide\addressmark[Tu],
        I. Ots\addressmark[Tu], W. Porod\address[Zue]{Inst. f\"ur Theor. 
Physik, Universit\"at Z\"urich, CH-8057 Z\"urich, Switzerland},
R.Saar\addressmark[Tu],
H. Uibo\addressmark[Tu] }
       
\begin{document}

\begin{abstract}
A future $e^+e^-$ Linear Collider has a large physics potential for the
discovery of new physics beyond the Standard Model
and precision studies of the Standard Model itself.
It is well suited to complement and extend 
the physics program of the LHC concerning
the precise determination of the underlying theory.
The use of polarized beams at a LC will be one of
the decisive tools.
\vspace{1pc}
\end{abstract}
\maketitle

\vspace*{-1.9cm}
\section{Introduction}

\vspace*{-.3cm}
A future Linear Collider (LC) designed with high luminosity of about
${\cal L}=3.4\cdot10^{34}$cm$^{-2}$s$^{-1}$ at $\sqrt{s}=500$~GeV
and ${\cal L}=5.8 \cdot 10^{34}$cm$^{-2}$s$^{-1}$ at $\sqrt{s}=800$~GeV
will provide precision studies in the
energy range between LEP and the TeV range \cite{TDR}.
Concerning the searches
for New Physics it complements and extends the physics program of the 
Hadron Colliders,
Tevatron and the future Large Hadron Collider 
(LHC)\footnote[1]{contact Georg Weiglein, see
http://www.ippp.dur.ac.uk/ $\tilde{}$ georg/lhclc/}.
The LC has large potential for the discovery of new particles  
and is unique concerning the precision analysis of
new physics (NP) as well as of the Standard Model (SM)
\cite{TDR}. An important tool of a LC
is the use of polarized beams \cite{Steiner}. 
Already
in the base line design it is foreseen to use the electron beams 
polarized to around $80\%$ via a strained photocathode
technology. In order to generate also polarized positrons 
the use of a helical undulator is favoured, producing
polarized photons which generate via pair production polarized positrons
of about 40\% (with full intensity of the $e^+$ beam) to 60\% 
(with 55\% intensity).

In the next sections we summarize the new physics searches 
at a LC starting from high precision studies of the SM as a motivation.

\vspace*{-.3cm}
\section{High precision analysis of the SM}

\vspace*{-.3cm}
One possibility to see hints for new physics, even if new particles
are not yet found, are electroweak precision tests with an
unprecedented accuracy.

\vspace*{-.2cm}
\subsection{Triple gauge couplings in $e^+e^-\to Z \gamma$}
The study of triple gauge couplings is very promising. 
In \cite{Ots} a
method has been worked out to derive complete analytical expressions
for the spin polarization and alignment of the $Z$ boson in
$e^+e^-\to Z \gamma$ including the contributions from $Z Z
\gamma$ and $Z \gamma\gamma$ anomalous couplings. 

\vspace*{-.2cm}
\subsection{Anomalous couplings in $e^+e^-\to~W^+W^-$}
Searches for deviations from the SM have already been studied,
 comparing polarized cross
sections in a general parametrization including anomalous gauge
couplings with expected precision measurements at the LC \cite{TDR}. 
It turns out
that for this purpose the polarization of the beams is very powerful:
e.g. the polarization of $P(e^-)=\pm 80\%$ (together with
$P(e^+)=\mp 60\%$) improves the sensitivity up to a factor 1.8 (2.5).

\vspace*{-.2cm}
\subsection{Transversely polarized beams in $e^+e^-\to W^+ W^-$}
A promising possibility to study electroweak symmetry breaking
is the use of transversely polarized $e^+e^-$ beams which projects out
$W^+_L W^-_L$ \cite{Karol}.  The asymmetry with respect
to the azimuthal angle of this process focusses on the $LL$ mode.
This asymmetry is very pronounced
at high energies reaching about $10\%$.
The advantage of this observable is that at high energies this asymmetry
peaks at larger angles and not
in beam direction where the analysis might be difficult.  One has
to note, however, that for the use of transverse
beams the polarization of 
both beams is needed. The effect does not occur if only one beam is polarized
since the cross section is given by:
\begin{eqnarray}
\sigma&=&(1-P_{e^-}^L P_{e^+}^L)\sigma_{unp}+(P_{e^-}^L-P_{e^+}^L)
\sigma_{pol}^L\nonumber
\\&&+
P_{e^-}^T P_{e^+}^T \sigma_{pol}^T.
\label{eq_trans}
\end{eqnarray}
\subsection{GigaZ}
A very spectacular method for probing the SM is foreseen as an upgrade
of a LC: the GigaZ option. Here 
$e^+e^-\to Z\to f \bar{f}$ is studied at the pole with very high luminosity
of about $10^9$ Z's within three month. With this option the
effective 
electroweak leptonic mixing angle can be measured via the left--right
asymmetry of this process with an unprecedented accuracy, see
Table~\ref{tab_prec} \cite{TDR}.
\subsection{Sensitivity to CP violation in the SM}
In \cite{Stahl} electroweak dipole form factors of the $\tau$ lepton
have been analysed
with regard to CP violation. 
CP--odd triple product correlations have
been studied and sensitivity bounds for the real and imaginary
parts of these form factors have been set. 
Using polarized beams a LC could be
sensitive to these form factors up to $O(10^{-19})$~ecm.

\vspace*{-.1cm}
\section{Revealing the structure of Susy}

\vspace*{-.3cm}
Supersymmetry 
is widely regarded as the best motivated extension of the SM.
However, since the SM particles and their Susy partners are not mass
degenerate, 
Susy has to be broken, which leads even for its minimal
version, the Minimal Supersymmetric Standard Model (MSSM), to about 100 free
parameters. In specific scenarios of Susy breaking one has much less 
parameters: 5 in mSUGRA, 4 in AMSB and 5 in GMSB.

In order to exactly pin down the structure of the underlying model it
is unavoidable to extract the parameters without assuming a particular
breaking scheme. The LC with its clear signatures is well suited to
determine the general parameters with high precision and to test
fundamental Susy assumptions as e.g. the equality of quantum numbers
or of couplings of the particles and their Susy partners.  It turns
out that the use of polarized beams plays a decisive role in this
context.

\vspace*{-.1cm}
\subsection{Stop mixing angle in $e^+e^-\to \tilde{t}_1\tilde{t}_1$}
As demonstrated 
in \ci{TDR} the mass and the mixing angle of $\tilde{t}$ can be extracted 
with high precision via the study of polarized cross sections
for light stop production. At a high luminosity LC and with $P(e^-)=80\%$ and 
$P(e^+)=60\%$ an accuracy of $\delta(m_{\tilde{t}_1})\approx 0.8$~GeV and
$\delta\cos\theta_{\tilde{t}}\approx 0.008$ could be reachable.

\vspace*{-.1cm}
\subsection{Quantum numbers in 
$e^+e^-\to~\tilde{e}^{+}_{L,R}\tilde{e}^-_{L,R}$} 
Susy transformations associate $e^-_{L,R}\leftrightarrow \tilde{e}^-_{L,R}$ 
and the antiparticles 
$e^+_{L,R}\leftrightarrow \tilde{e}^+_{R,L}$. In order to prove
this association of scalar particles to chiral quantum numbers the use
of polarized beams is necessary \cite{Bloechi}. The process occurs via
$\gamma$ and $Z$ exchange in the s--channel and via $\tilde{\chi}^0_i$
exchange in the t--channel. Only in the t--channel the SM particle is
directly coupled to its scalar partner and in order to test the
association of quantum numbers one has to projects out the t--channel 
exchange.

\begin{figure}[htb]
\vspace{75pt}
\begin{picture}(63,53)
\put(15,-2){\mbox{\epsfig{figure=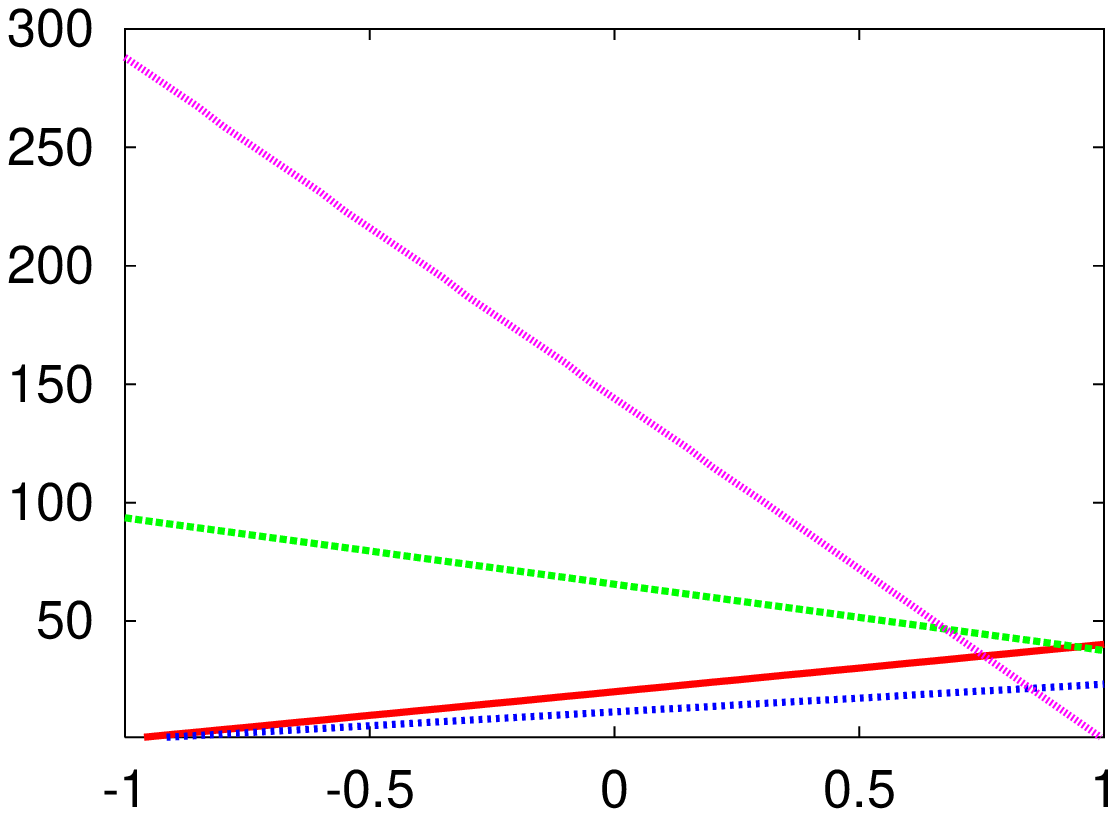,height=5cm,width=6.5cm}}}
\put(-2,125){\makebox(0,0)[bl]{{$\sigma$[fb]}}}
\put(100,100){\makebox(0,0)[br]{{$\tilde{e}^-_{L} \tilde{e}^+_{R}$}}}
\put(100,50){\makebox(0,0)[br]{{$\tilde{e}^-_{R} \tilde{e}^+_{R}$}}}
\put(215,30){\makebox(0,0)[br]{{$\tilde{e}^-_{R} \tilde{e}^+_{L}$}}}
\put(215,15){\makebox(0,0)[br]{{$\tilde{e}^-_{L} \tilde{e}^+_{L}$}}}
\put(210,-5){\makebox(0,0)[br]{{$P_{e^+}$}}}
\put(180,100){\makebox(0,0)[br]{{$P_{e^-}=-80\%$}}}
\put(180,115){\makebox(0,0)[br]{{$\sqrt{s}=400$~GeV}}}
\end{picture}
\vspace{-25pt}
\caption{Test of selectron quantum numbers in
$e^+e^-\to\tilde{e}_{L,R}^+\tilde{e}_{L,R}^-$ with fixed electron
polarization $P(e^-)=-80\%$ and variable positron polarization
$P(e^+)$. For  $P(e^-)=-80\%$ and $P(e^+)<0$ both pairs 
$\tilde{e}_L^- \tilde{e}_R^+$ and 
$\tilde{e}^-_R \tilde{e}^+_R$ still contribute. For $P(e^+)=-60\%$
$\tilde{e}_L^- \tilde{e}_R^+$ dominates by more than a factor 3  
\cite{Bloechi}.}
\label{fig_sel}
\vspace{-25pt}
\end{figure}

With completely polarized 
$e^-_L e^+_L$ only $\tilde{e}^-_L \tilde{e}^+_R$ contributes. Due to
their $L,R$ coupling character $\tilde{e}_L$, $\tilde{e}_R$ can be
discriminated via their decays and can be identified via their charge. One
has to note that a polarized $e^+$ beam is necessary. Even completely
polarized $e^-$ would not be sufficient, since otherwise the s-channel 
exchange could
not be switched off.  In reality partially polarized beams
of $P_{e^-}=-80\%$ and $P_{e^+}=-60\%$ can
be sufficient to probe this association. With this polarization
$\tilde{e}^-_L\tilde{e}^+_R$ dominates e.g. by a factor
of 3, Fig.~\ref{fig_sel}.
\subsection{Gaugino/higgsino sector}
In \cite{CKMZ} and references therein a strategy has been worked out
to determine accurately 
the MSSM parameters $M_1$, $\Phi_{M_1}$, $M_2$, $\mu$,
$\Phi_{\mu}$ and moderate $\tan\beta$ via the study of polarized rates
in $e^+ e^-\to \tilde{\chi}^{\pm}_i\tilde{\chi}^{\mp}_j$ and
$e^+e^-\to\tilde{\chi}^0_i\tilde{\chi}^0_j$. 
Even if only $\tilde{\chi}^+_1\tilde{\chi}^-_1$, 
$\tilde{\chi}^0_1\tilde{\chi}^0_2$ were accessible, it 
would be sufficient for determining the MSSM parameters \cite{CKMZ}.  
With light chargino
pair production and polarized initial beams one can unambiguously determine
the mixing angle but $M_2$, $\mu$, $\tan\beta$ still depend
on $m_{\tilde{\chi}^{\pm}_2}$. If one furthermore includes now the light
neutralino system and studies the polarized cross
sections $e^+ e^-\to \tilde{\chi}^0_1 \tilde{\chi}^0_2$ in the complex
$M_1$ plane, one can simultaneously
determine the parameter $M_1$ and predict
$m_{\tilde{\chi}^{\pm}_2}$.

Once the parameters are determined one can efficiently test 
whether the gauge couplings $g_{Bee}$ and $g_{Wee}$
are identical to the Yukawa couplings
$g_{\tilde{B}e\tilde{e}}$ and $g_{\tilde{W} e \tilde{e}}$, respectively, by
studying the polarized cross sections with a variable ratio of
$g_{Bee}/g_{\tilde{B}e\tilde{e}}$ and $g_{Wee}/g_{\tilde{W} e \tilde{e}}$
and comparing it with experimental values \cite{CKMZ}.

\vspace*{-.2cm}
\subsection{Case of high $\tan\beta$: $\tau$ polarization}
In case of high $\tan\beta>10$ the chargino and neutralino sector is
insensitive to this parameter. However, one could then  
determine $\tan\beta$ from another sector 
whose particles are relatively
light in many scenarios: the $\tilde{\tau}$ sector.

\begin{figure}[htb]
\epsfig{figure=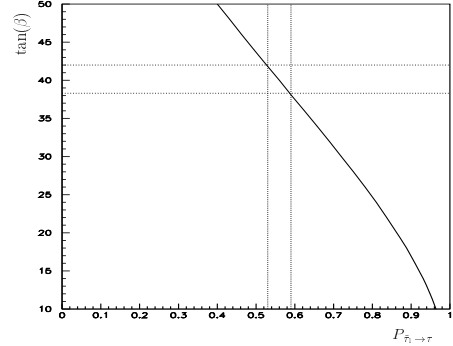, angle=0,width=6.5cm,height=5cm,clip=true}
\vspace{-32pt}
\caption{Measuring the $\tau$ polarization 
$P_{\tilde{\tau}_1\to\tau}$ from the decay 
$\tilde{\tau}_1\to \tau\tilde{\chi}^0_1$ for an already measured mixing angle 
$\cos\theta_{\tilde{\tau}}$ leads in an example of \cite{Stau}
to an accurate determination of high $\tan\beta$: $\tan\beta=40\pm 2$.}
\label{fig_stau}
\vspace{-15pt}
\end{figure}

The polarization of $\tau$'s from $\tilde{\tau}_i\to\tau\tilde{\chi}^0_1$ 
is sensitive to $\tan\beta$. In case of a sufficient higgsino admixture
in the $\tilde{\chi}^0_1$ it is even possible to determine high
$\tan\beta$ as well as $A_{\tau}$, without any assumptions on the
Susy breaking mechanism \cite{Stau}: after 
determining the $\tilde{\tau}$ mixing
angle via the polarization asymmetry 
$A_{pol}=(\sigma_L-\sigma_R)/(\sigma_L+\sigma_R)$
in $e^+e^-\to\tilde{\tau}_1\tilde{\tau}_1$,
one can determine $\tan\beta$ from
the $\tau$ polarization in the decay $\tilde{\tau}\to \tau
\tilde{\chi}^0_1$, see Fig.\ref{fig_stau}. 
 Even for high $\tan\beta\approx 40$ one can reach
an accuracy of about $10\%$.  
Measuring now $m_{\tilde{\tau}_2}$ leads to the determination of
$A_{\tau}$.
\subsection{CP--phases in the $\tilde{\tau}$ sector}
In the $\tilde{\tau_{i}}$ system CP--violating phases can occur in the
parameters $A_{\tau}$, $\mu$ and $M_1$. These phases have a
decisive influence on masses, couplings and various CP--violating observables.
  The influence of the phases could be quite strong
even on CP--conserving decay branching ratios (BR's) 
and hence one should take it into account when determining the MSSM
parameters at the LHC and LC. In this context the BR's also might be a
sensitive probe for the CP--phases \cite{Hidaka}.
\subsection{Extended Susy models}
In case of R--parity violating Susy non--standard couplings could
occur which produce a scalar particle in the s--channel:
$e^+e^-\to\tilde{\nu}\to e^+e^-$. The process gives a significant signal
over the background. Since it requires both left--handed $e^-$ and
$e^+$ beams, it can be easily analysed and identified
by the use of beam polarization
(\cite{Steiner} and references therein): here 
simultaneously polarized beams
enhance the signal by about a factor of 10.\\
\hspace*{.0cm}Introducing in the MSSM an additional Higgs singlet
leads to the (M+1)SSM with one additional neutralino. 
Since the mass spectra of
the four light neutralinos could be similar to those in the MSSM a
distinction between the models might be difficult via spectra and rates alone.
However, polarization effects might then
indicate the different coupling structure in the (M+1)SSM \cite{Hesselbach}.

\vspace*{-.2cm}
\section{Other kind of New Physics}

\vspace*{-.3cm}
Another approach to resolve the hierachy problem is the introduction
of large extra dimensions. At a LC the process $e^+e^-\to \gamma G$ 
is promising and it has been worked out that running on two
different $\sqrt{s}$ one can determine the number of extra dimensions
\cite{TDR}. The use of polarized beams in this context enlarges on
one hand the sensitivity to the new scale $M_{*}$ and suppresses
on the other hand
the main background $e^+e^-\to \nu \nu \gamma$ significantly. 
The ratio $S/\sqrt{B}$ is enhanced by a factor of about 2.1 (4.4) if
$P(e^-)=+80\%$ (and $P(e^+)=-60\%$) is used.

\vspace*{-.2cm}
\section{Summary}

\vspace*{-.3cm}
It has been shown that a LC in the TeV range is well suited for new
physics searches and in particular for its precise determination.  The
use of polarized beams plays a decisive role\footnote[2]{for updates see
POWER group (POlarization at Work in Energetic Reactions), forum for
machine physicists, experimentalists, theorists: http://www.desy.de/ $\tilde{}$
gudrid/POWER/}.  In this context the use of simultaneously polarized
$e^-$, $e^+$ beams has several advantages: enhancing considerably the accuracy
for SM precision tests at GigaZ, providing higher sensitivity to
non--standard couplings and supplying efficient tools to identify quantum
numbers of the new particles and to determine 
the underlying parameters of the theory.

\vspace{.3cm}
G.M.--P. was partially supported by
the Graduiertenkolleg `Zuk\"unftige Entwicklungen in der Teilchenphysik' of
the University of Hamburg, Project No. GRK 602/1.
This work was also supported by the EU TMR Network Contract No.
HPRN-CT-2000-00149. The group from IP, University of Tartu was partly 
supported by ESF grant 4510.

\vspace{-.2cm}
\begin{table}
\caption{Ew. precision parameters at different 
colliders \cite{gigaz} \label{tab_prec}}
{\footnotesize\hspace*{-.4cm}
\begin{tabular}{l|cccc}
Accuracy & LEP/Tev. & Tev./LHC & LC & GigaZ \\ \hline
$\delta M_W$ & 34 MeV & 15 MeV & 15 MeV & 6 MeV\\
$\delta \sin^2\theta_{eff}$ & 0.00017 & 0.00017 & 0.00017 & 0.00001\\
$\delta m_t$ & 5 GeV & 2 GeV & 0.2 GeV & 0.2 GeV \\
$\delta m_h$ & -- & 0.2 GeV & 0.05 GeV & 0.05 GeV\\[-.8cm]
\end{tabular}
}
\end{table}


\end{document}